\documentstyle[12pt]{article}

\begin{document}
\input{psfig.sty}

\newcommand{\drawsquare}[2]{\hbox{%
\rule{#2pt}{#1pt}\hskip-#2pt
\rule{#1pt}{#2pt}\hskip-#1pt
\rule[#1pt]{#1pt}{#2pt}}\rule[#1pt]{#2pt}{#2pt}\hskip-#2pt
\rule{#2pt}{#1pt}}

\newcommand{\Yfund}{\raisebox{-.5pt}{\drawsquare{6.5}{0.4}}}
\newcommand{\Ysymm}{\raisebox{-.5pt}{\drawsquare{6.5}{0.4}}\hskip-0.4pt%
        \raisebox{-.5pt}{\drawsquare{6.5}{0.4}}}
\newcommand{\Ythrees}{\raisebox{-.5pt}{\drawsquare{6.5}{0.4}}\hskip-0.4pt%
          \raisebox{-.5pt}{\drawsquare{6.5}{0.4}}\hskip-0.4pt%
          \raisebox{-.5pt}{\drawsquare{6.5}{0.4}}}
\newcommand{\Yfours}{\raisebox{-.5pt}{\drawsquare{6.5}{0.4}}\hskip-0.4pt%
          \raisebox{-.5pt}{\drawsquare{6.5}{0.4}}\hskip-0.4pt%
          \raisebox{-.5pt}{\drawsquare{6.5}{0.4}}\hskip-0.4pt%
          \raisebox{-.5pt}{\drawsquare{6.5}{0.4}}}
\newcommand{\Yasymm}{\raisebox{-3.5pt}{\drawsquare{6.5}{0.4}}\hskip-6.9pt%
        \raisebox{3pt}{\drawsquare{6.5}{0.4}}}
\newcommand{\Ythreea}{\raisebox{-3.5pt}{\drawsquare{6.5}{0.4}}\hskip-6.9pt%
        \raisebox{3pt}{\drawsquare{6.5}{0.4}}\hskip-6.9pt
        \raisebox{9.5pt}{\drawsquare{6.5}{0.4}}}
\newcommand{\Yfoura}{\raisebox{-3.5pt}{\drawsquare{6.5}{0.4}}\hskip-6.9pt%
        \raisebox{3pt}{\drawsquare{6.5}{0.4}}\hskip-6.9pt
        \raisebox{9.5pt}{\drawsquare{6.5}{0.4}}\hskip-6.9pt
        \raisebox{16pt}{\drawsquare{6.5}{0.4}}}
\newcommand{\Yadjoint}{\raisebox{-3.5pt}{\drawsquare{6.5}{0.4}}\hskip-6.9pt%
        \raisebox{3pt}{\drawsquare{6.5}{0.4}}\hskip-0.4pt
        \raisebox{3pt}{\drawsquare{6.5}{0.4}}}
\newcommand{\Ysquare}{\raisebox{-3.5pt}{\drawsquare{6.5}{0.4}}\hskip-0.4pt%
        \raisebox{-3.5pt}{\drawsquare{6.5}{0.4}}\hskip-13.4pt%
        \raisebox{3pt}{\drawsquare{6.5}{0.4}}\hskip-0.4pt%
        \raisebox{3pt}{\drawsquare{6.5}{0.4}}}
\newcommand{\Yflavor}{\Yfund + \overline{\Yfund}} 
\newcommand{\Yoneoone}{\raisebox{-3.5pt}{\drawsquare{6.5}{0.4}}\hskip-6.9pt%
        \raisebox{3pt}{\drawsquare{6.5}{0.4}}\hskip-6.9pt%
        \raisebox{9.5pt}{\drawsquare{6.5}{0.4}}\hskip-0.4pt%
        \raisebox{9.5pt}{\drawsquare{6.5}{0.4}}}%

\begin{flushright}
\baselineskip=12pt
UPR-988-T \\
\end{flushright}

\begin{center}
\vglue 1.5cm
{\Large\bf Deconstruction of Gauge Symmetry Breaking by Discrete Symmetry
and $G^N$ Unification}
\vglue 2.0cm
{\Large Tianjun Li~\footnote{Current Address: 
School of Natural Sciences,  Institute for Advanced Study, Einstein Drive, 
Princeton, NJ 08540.
E-mail: tli@sns.ias.edu. Phone Number: (609) 734-8024. Fax Number:  (609) 
951-4489.}
and Tao Liu~\footnote{E-mail: liutao@sas.upenn.edu.}}
\vglue 1cm
{ Department of Physics and Astronomy \\
University of Pennsylvania, Philadelphia, PA 19104-6396 \\  
U.  S.  A.}
\end{center}

\vglue 1.5cm
\begin{abstract}
We deconstruct the non-supersymmetric
 $SU(5)$ breaking by discrete symmetry
 on the space-time $M^4\times S^1$
 and $M^4\times S^1/(Z_2\times Z_2')$ in the
 Higgs mechanism deconstruction
scenario.
And we explain the subtle point on how to exactly
match the continuum results with the latticized results
on the quotient space $S^1/Z_2$ and $S^1/(Z_2\times Z_2')$.
We also propose an effective deconstruction scenario
and discuss the gauge symmetry breaking by the discrete symmetry
on theory space in this approach.
As an application,
we suggest the $G^N$ unification where $G^N$ is
broken down to $SU(3)\times SU(2)\times U(1)^{n-3}$
by the bifundamental link fields and the doublet-triplet
splitting can be achieved.
\\[1ex]
\\[1ex]

\end{abstract}

\vspace{0.5cm}
\begin{flushleft}
\baselineskip=12pt
April 2002\\
\end{flushleft}
\newpage
\setcounter{page}{1}
\pagestyle{plain}
\baselineskip=14pt

\section{Introduction}
Grand Unified Theory (GUT) gives us a simple
 and elegant understanding of the quantum numbers of quarks and leptons,
and the success of gauge coupling unification in the Minimal 
Supersymmetric
Standard Model strongly supports
 this idea. Although the Grand Unified Theory at high energy scale has
been widely accepted, there are some problems in GUT: 
 the grand unified
gauge symmetry breaking mechanism, the doublet-triplet splitting problem, 
and
the proton decay problem, etc.

Recently, a new scenario proposed to address above questions in GUT has
 been discussed extensively~\cite{SBPL, JUN}.
 The key point is that the GUT
gauge symmetry exists in 5 or higher dimensions and is broken down to the
4-dimensional 
$N=1$ supersymmetric Standard Model like gauge symmetry for
 the zero modes due to the 
discrete symmetries in the brane neighborhoods
or on the extra space manifolds, which
become non-trivial constraints on the multiplets and gauge generators in 
GUT~\cite{JUN}. 
The attractive models have been constructed explicitly, where
the supersymmetric 5-dimensional and 6-dimensional 
GUT models are broken down to
the 4-dimensional $N=1$
 supersymmetric $SU(3)\times SU(2) \times U(1)^{n-3}$
model, where $n$ is the rank of GUT group, through the 
compactification on various orbifolds and manifolds. 
The GUT gauge symmetry breaking and doublet-triplet
splitting problems have been solved neatly by the
discrete symmetry projections.
Other interesting phenomenology, like $\mu$ problems, gauge coupling
unifications, non-supersymmetric GUT, gauge-Higgs unification,
proton decay, etc, have also been discussed~\cite{SBPL, JUN}. 

On the other hand, deconstrution was proposed about one year 
ago~\cite{ACG}.
Deconstruction is interesting because
 it provides a UV completion of the higher dimensional theories.
A lot of phenomenological and formal issues in deconstruction
scenarios have been discussed. These include
 the extensions of the
Standard Model, gaugino mediated supersymmetry breaking,
low energy unification, GUT breaking, electroweak
symmetry breaking, anomaly inflow, the description of little
string theories in terms of gauge theory, a single 
gauge group description of extra dimensions in the large
number of color limit, models with noncommutative geometry,
warped background geometry, topological objects, 
Seiberg-Witten curves, and even the deconstruction of time and
gravity, etc~\cite{DCP, CKT, CMW, NW, EW}.
By the way, the arrays of gauge theories, where 
an infinite number of gauge theories are linked by scalars,
were discussed previously~\cite{MBH}.

In this paper, we would like to discuss the 
deconstruction of  gauge symmetry breaking by the
discrete symmetry on extra space manifold.
If we know how to deconstruct those higher dimensional
theories, we can have a lot of good features in deonstruction
scenarios, for
example, the gauge symmetry breaking, the doublet-triplet
splitting, suppressing the proton decay, the gauge-Higgs unification
(In the deconstruction language, bifundamental fields and $SU(2)_L$
Higgs unification.).
For simplicity, we do not consider supersymmetry.
First, we consider the Higgs mechanism deconstruction scenario in which
the gauge bosons obtain the masses via the VEVs of the
bifundamental link Higgs fields.
We shall deconstruct the non-supersymmetric
 $SU(5)$ breaking by the discrete symmetry on the space-time $M^4\times 
S^1$
 and $M^4\times S^1/(Z_2\times Z_2')$ where
$M^4$ is the 4-dimensional Minkowski space-time.
We also explain the subtle point on how to exactly
match the continuum results with the latticized results
on the quotient space $S^1/Z_2$ and $S^1/(Z_2\times Z_2')$.
In addition, it seems to us that
the Higgs mechanism deconstructions of the 
gauge symmetry breaking by discrete symmetry 
on  the space-time $M^4\times S^1$
 and $M^4\times S^1/(Z_2\times Z_2')$
 might not be the real deconstructions. The key point is that,
the bifundamental field $U_i$, which is the Schwinger line integral
along the fifth dimension,
should be considered as the gauge field $A_5$.
However, the mass spectrum and 
the 5-dimensional wave functions for the
KK modes of $A_5$ 
 can not match those for $U_i$ 
in the Higgs mechanism deconstruction scenario by counting
the massless modes\footnote{One might consider the axial gauge $A_5=0$ in
the 5-dimensional theory. However, only part of the scalars from $U_i$ 
fields
are eaten by the massive gauge bosons after the gauge symmetry breaking, 
and
most of the physical scalars from $U_i$ fields can obtain masses via
Higgs mechanism. Therefore, the 5-dimensional gauge field $A_5$ in the 
axial gauge ($A_5=0$)
can not match the bifundamental fields $U_i$ in the Higgs mechanism 
deconstruction scenario, too.
}.
Therefore, we propose the effective deconstruction
scenario where we add the mass term for each field by hand which
comes from the latticization of the kinetic term of
that field along the fifth dimension. In the effective deconstruction 
scenario,
similar to the gauge symmetry breaking by the discrete
symmetry on extra space manifold,
we can define the discrete symmetry on theory space and
discuss the gauge symmetry breaking.
Moreover, we show that the continuum results match the effective
deconstruction results exactly.
As an application,
we discuss the $G^N$ unification where $G^N$ is
broken down to $SU(3)\times SU(2)\times U(1)^{n-3}$
($n$ is the rank of group $G$)
by the bifundamental link fields and the doublet-triplet
splitting can be achieved. 
With the general $G^N$ unification, we wish we can solve the tough 
problems
in the traditional 4-dimensional
GUT models. 

Let us explain our terminology. The exactly match between the 
$n-th$ KK mode of a bulk field in the continuum case and
the corresponding state in the deconstruction case
 means that the
$n-th$ mass eigenvalue and eigenvector of the field in the deconstruction
scenario are the same as the mass and 5-dimensional wave function
of the $n-th$ KK mode of the corresponding bulk field when $N > > n$.

\section{Higgs Mechanism Deconstruction of
$SU(5)$ Breaking on $M^4\times S^1$ by Wilson Line}

In this section, we would like to deconstruct
 the non-supersymmetric
$SU(5)$ breaking on the space-time $M^4\times S^1$ by Wilson
line in the Higgs mechanism deconstruction scenario.
 And we want to point out that one can discuss any other
GUT groups similarly because the fundamental group of $S^1$
is $Z$, {\it i. e.}, $\pi_1 (S^1)=Z$~\cite{JUN}.

\subsection{$SU(5)$ Breaking on $M^4\times S^1$ by Wilson Line}
Let us consider 
the 5-dimensional space-time which can be factorized into 
a product of the 
ordinary 4-dimensional Minkowski space-time $M^4$ and the circle 
$S^1$. The corresponding coordinates are $x^{\mu}$, 
($\mu = 0, 1, 2, 3$),
$y\equiv x^5$, and the radius for the fifth dimension is $R$.
The gauge fields are denoted as $A_M (x^{\mu}, y)$ where 
$M=0, 1, 2, 3, 5$.
Because $Z_2 \subset \pi_1 (S^1)$, we can define the $Z_2$ parity
operator $P$ for a generic bulk multiplet $\Phi(x^{\mu}, y)$ 
\begin{eqnarray}
\Phi(x^{\mu},y)&\to \Phi(x^{\mu},y+ 2\pi R)=
\eta_{\Phi} P^{\l_{\Phi}}\Phi(x^{\mu},y)
(P^{-1})^{m_{\Phi}}~,~\,
\end{eqnarray}
where $\eta_{\Phi} = \pm 1$ and $P^2=1$. 
By the way, if the gauge group $G$ is
$SU(5)$, for a 5-plet $\Phi$ in the fundamental representation,
 $\l_{\Phi}=1$ and $m_{\Phi}=0$, and
 for a 24-plet
$\Phi$ in the adjoint representation, $\l_{\Phi}=1$ and $m_{\Phi}=1$.

Denoting the field $\phi$ with parity $P$=$\pm$ by $\phi_{\pm}$, 
we obtain the KK mode expansions
\begin{eqnarray}
  \phi_{+} (x^\mu, y) &=& 
      \sum_{n=-\infty}^{+\infty} \phi_{+}^n (x^{\mu}) e^{i{{ny}\over 
R}}~,~\,
\end{eqnarray}
\begin{eqnarray}
  \phi_{-} (x^\mu, y) &=& 
      \sum_{n=-\infty}^{+\infty} \phi_{-}^n (x^{\mu}) 
e^{i{{(n+1/2)y}\over R}}~.~\,
\end{eqnarray}

Now, let us discuss the $SU(5)$ breaking. Under parity $P$,
the gauge fields $A_M$ transform as 
\begin{eqnarray}
  A_M (x^\mu, y+2 \pi R) &=& P A_M (x^\mu, y) P^{-1}
~.~\,
\end{eqnarray}

And we choose the following matrix representation
for parity operator $P$, which is expressed in the
adjoint representation of $SU(5)$
\begin{equation}
P={\rm diag}(-1, -1, -1, +1, +1)
 ~.~\,
\end{equation}
So, upon the $P$ parity,
 the gauge generators $T^A$ where A=1, 2, ..., 24 for $SU(5)$
are separated into two sets: $T^a$ are the gauge generators for
the Standard Model gauge group, and $T^{\hat a}$ are the other broken
gauge generators 
\begin{equation}
P~T^a~P^{-1}= T^a ~,~ P~T^{\hat a}~P^{-1}= - T^{\hat a}
~.~\,
\end{equation}
And the masses for $A^a_M$ and $A^{\hat a}_M$
are $n/R$ and $(n+1/2)/R$, respectively.
In addition, if we add a pair of
 Higgs 5-plets $H_u$ and $H_d$ in the bulk,
then for each 5-plet,
the doublet mass is  $n/R$ and the triplet mass is $(n+1/2)/R$
if $\eta_{H_u}=\eta_{H_d}=+1$.
In short, for the zero modes, the gauge group $SU(5)$ is
broken down to $SU(3)\times SU(2)\times U(1)$, and
we can solve the doublet-triplet splitting problem.
The parities and masses of the fields
in the $SU(5)$ gauge and Higgs multiplets are given in Table 1.

\renewcommand{\arraystretch}{1.4}
\begin{table}[t]
\caption{Parity assignment and masses of the fields in the $SU(5)$
 gauge and Higgs multiplets for the model with $SU(5)$ breaking
by Wilson line.
The index $a$ labels the unbroken $SU(3)\times SU(2) \times U(1)$ 
gauge generators, while $\hat a$
labels the other broken SU(5) gauge generators.
The indices $D$ and $T$ are for doublet and triplet, respectively. 
\label{tab:chiral}}
\vspace{0.4cm}
\begin{center}
\begin{tabular}{|c|c|c|}
\hline        
$P$ & field & mass ($n=0, \pm1, \pm2, ...$)\\ 
\hline
$+$ &  $A^a_{\mu}$, $A^a_5$, $H^D_u$, $H^D_d$ & $n/R$ \\
\hline
$-$ &  $A^{\hat{a}}_{\mu}$, $A^{\hat{a}}_5$  $H^T_u$, $H^T_d$ & 
$(n+1/2)/R$ \\
\hline
\end{tabular}
\end{center}
\end{table}

\subsection{Higgs Mechanism Deconstruction}
We consider the $SU(5)^{N+1}$
 gauge theory with bifundamental
fields $U_i$ as follows

\begin{equation}
\label{wilson}
\begin{array}{c|cccccc}
      & SU(5)_0 & SU(5)_1 & SU(5)_2 & \cdots & SU(5)_{N-1} & SU(5)_N \\
\hline
{\vrule height 15pt depth 5pt width 0pt}
 U_0       & \Yfund            & \overline{\Yfund}  & 1     & \cdots & 1 
&1 \\
  U_1       & 1       & \Yfund & \overline{\Yfund}   & \cdots & 1 & 1 \\
  \vdots    & \vdots & \vdots & \ddots & \vdots & \vdots \\
  U_{N-1}   & 1 & 1 & 1& \cdots & \Yfund & \overline{\Yfund} \\
U_N & \overline{\Yfund}  & 1& 1 & \cdots & 1 & \Yfund \\
\end{array}. \nonumber
\end{equation}

And the effective action is
\begin{eqnarray}
 S&=&\int d^4x \sum_{i=0}^N \left(-{1\over {4 g^2}} Tr F_i^2 + 
Tr[(D_{\mu} U_i)^{\dagger} D^{\mu} U_i]+...\right)
~,~\,
\end{eqnarray}
where the covariant derivative is 
$D^{\mu} U_i\equiv \partial_{\mu} U_i-i A_{\mu}^i U_i+i U_i A_{\mu}^{i+1}$
and the dots represent the higher dimensional operators that are
irrelevant at low energies.

The bifundamental fields $U_i$ obtain the vacuum expectation values (VEV)
either from a suitable renormalizable potential or from
some strong interactions. In order to obtain the deconstruction of
the Wilson line gauge symmetry breaking, we choose the following
VEVs for $U_i$
\begin{eqnarray}
 < U_i>
&=& {\rm diag} (v/\sqrt 2, v/\sqrt 2, v/\sqrt 2, v/\sqrt 2, v/\sqrt 2),
~{\rm for }~i=0, 1, 2, ..., N-1 
~,~\,
\end{eqnarray}
\begin{eqnarray}
 < U_N>
&=& {\rm diag} (-v/\sqrt 2, -v/\sqrt 2, -v/\sqrt 2, v/\sqrt 2, v/\sqrt 2)
~.~\,
\end{eqnarray}
We would like to explain the VEV for $U_N$, which is different from
the VEV for $U_i$ where $i=0, 1, 2, ..., N-1$. In general, we can take
that the VEV for $U_N$ is the same as that for $U_i$. And the
masses for gauge bosons are given by the mass terms  
${1\over 2} \sum_{i=0}^{N} g^2 v^2 (A_{\mu}^{\beta (i+1)}-
A_{\mu}^{\beta i})^2$, where $A_{\mu}^{\beta (N+1)} = A_{\mu}^{\beta 0}$
for $\beta =a$ and $A_{\mu}^{\beta (N+1)} = - A_{\mu}^{\beta 0}$
for $\beta =\hat a$ due to the Wilson line gauge symmetry breaking
(See Table 1). However, in the Higgs mechanism deconstruction
scenario, we take $A_{\mu}^{\beta (N+1)} = A_{\mu}^{\beta 0}$
for $\beta =a$ and  $\beta =\hat a$, so, the last term in above mass terms
is $ {1\over 2}  g^2 v^2 (A_{\mu}^{\beta 0} -
A_{\mu}^{\beta N})^2$ for $\beta =a$, and
$ {1\over 2}  g^2 v^2 (A_{\mu}^{\beta 0} + 
A_{\mu}^{\beta N})^2$ for $\beta =\hat a$.
This correct mass term can be obtained by choosing the suitable 
VEV for $U_N$, which is the Eq. (10) in our model. By the way,
we emphasize that the VEV for $U_N$ in Eq. (10) is similar to the
matrix representation
for parity operator $P$ in Eq. (5), which breaks the $SU(5)$ gauge
symmerty.

Thus, the $(N+1)\times (N+1)$ mass matrix for the Standard Model
gauge boson or
the column vector 
$(A_{\mu}^{a 0}, A_{\mu}^{a 1}, A_{\mu}^{a 2},..., A_{\mu}^{a N})$
is 
\begin{equation}
\label{SM22}
M_{SM}^2 = g^2 v^2\left(
\begin{array}{ccccc}
2&-1&0&\cdots&-1 \\
-1&2&-1& \cdots&0 \\
0&-1 &2 &\cdots&0 \\
& & & \cdots & \\
-1&0&\cdots&-1&2
\end{array} \right).
\end{equation}
The mass spectrum or eigenvalue is 
\begin{eqnarray}
M_n^2= 4 g^2 v^2 \sin^2({{n\pi}\over {N+1}})
~,~\,
\end{eqnarray}
where $-N/2 \leq n \leq N/2$ or $n=0, 1, 2, ..., N$.
And the corresponding $n-th$ eigenvector is
\begin{eqnarray}
\alpha_n=(\alpha_{n}^0, \alpha_n^1, ..., \alpha_n^N)
~,~\,
\end{eqnarray}
where
\begin{eqnarray}
 \alpha_n^j &=& {1\over {\sqrt {N+1}}} 
{\rm exp}(i{{2 j n\pi}\over {N+1}}), ~j=0, 1, ..., N
~.~\,
\end{eqnarray}
Noticing that $R=(N+1)/(gv)$, we obtain that the $n-th$ eigenvector 
matches
the 5-dimensional wave function ($e^{i{{ny}\over R}}$) of
the $n-th$ KK mode for $A_{\mu}^a$ in the last subsection,
and for $n<< N$, the $n-th$ mass (eigenvalue) matches the mass of the
$n-th$ KK mode for $A_{\mu}^a$. So, they exactly match.

And the  mass matrix for the non-Standard Model gauge boson or
the column vector 
$(A_{\mu}^{\hat{a} 0}, 
A_{\mu}^{\hat{a} 1}, A_{\mu}^{\hat{a} 2},..., A_{\mu}^{\hat{a} N})$
is 
\begin{equation}
\label{NSM22}
M_{NSM}^2 = g^2 v^2\left(
\begin{array}{ccccc}
2&-1&0&\cdots&+1 \\
-1&2&-1& \cdots&0 \\
0&-1 &2 &\cdots&0 \\
& & & \cdots & \\
+1&0&\cdots&-1&2
\end{array} \right).
\end{equation}

The mass spectrum or eigenvalue is 
\begin{eqnarray}
M_n^2= 4 g^2 v^2 \sin^2({{(n+1/2)\pi}\over {N+1}})
~,~\,
\end{eqnarray}
where $-N/2 \leq n \leq N/2$ or $n=0, 1, 2, ..., N$.
And the corresponding $n-th$ eigenvector is
\begin{eqnarray}
\alpha_n=(\alpha_{n}^0, \alpha_n^1, ..., \alpha_n^N)
~,~\,
\end{eqnarray}
where
\begin{eqnarray}
 \alpha_n^j &=& {1\over {\sqrt {N+1}}} 
{\rm exp}(i{{j (2n+1)\pi} \over {N+1}}), ~j=0, 1, ..., N
~.~\,
\end{eqnarray}
Noticing that $R=(N+1)/(gv)$, we obtain that the $n-th$ eigenvector 
matches
the 5-dimensional wave function ($e^{i{{(n+1/2)y}\over R}}$) of 
the $n-th$ KK mode for $A_{\mu}^{\hat {a}}$ in the last subsection,
and for $n<< N$, the $n-th$ mass (eigenvalue) matches the mass of the
$n-th$ KK mode for $A_{\mu}^{\hat {a}}$. Thus, they exactly match, and
we want to emphasize that there is no massless mode for $A_{\mu}^{\hat{a} 
i}$.

In addition, we can discuss the deconstruction of Higgs fields $H_u$ and 
$H_d$.
We assume that under the $i-th$ gauge group $SU(5)_i$, there is a pair of 
Higgs
5-plets $H_u^i$ and $H_d^i$ where $i=0, 1, 2, ..., N$. And we
consider the following potential
\begin{eqnarray}
V&=& 2 g^2 \sum_{i=0}^{N} (|U_i (H_u^i)^{\dagger}|^2 + |H_u^{i+1} U_i|^2 +
|U_i H_d^i|^2 + |(H_d^{i+1})^{\dagger} U_i|^2  )
\nonumber\\&& 
-\sqrt 2 g^2 v \sum_{i=0}^N (H_u^{i+1} U_i (H_u^i)^{\dagger} + 
(H_d^{i+1})^{\dagger} U_i H_d^i + H. C.)
~,~\,
\end{eqnarray}
where for simplicity, we define $H_u^{N+1} \equiv H_u^0$ and $H_d^{N+1} 
\equiv H_d^0$.
The above potential for a pair of Higgs
5-plets $H_u^i$ and $H_d^i$ comes from the deconstruction of $(D_5 
H_u)^{\dagger} D_5 H_u $ 
and $(D_5 H_d)^{\dagger} D_5 H_d $ in 5-dimensional theory, 
and the coefficients are determined by the normalization
which is compatible with that of gauge fields.
In short, we obtain that the mass matrix for the doublet $H_u^{iD}$ or 
$H_d^{iD}$  is the same as that
for the Standard Model gauge boson in Eq. (\ref{SM22}), and the 
mass matrix for the triplet $H_u^{iT}$ or $H_d^{iT}$  is the same as that
for the non-Standard Model gauge boson in Eq. (\ref{NSM22}). Therefore, 
similar to the discussions
for $A_{\mu}^a$ and $A_{\mu}^{\hat a}$, the deconstruction results match 
the
 continuum results in Table 1. And then, we solve the doublet-triplet 
splitting problem.

Furthermore, by counting the number of massless modes,
 we can prove that the bifundamental fields $U_i$ can not match
the gauge fields $A_5$ because $U_i$s are Higgs fields. The correct
deconstruction of the $A_5$ fields should have 12 massless modes, but, the
$U_i$s field will have at least $12(2N+1)$ massless modes that are the
Goldstone bosons and give 
 masses to the longitudinal components of massive gauge bosons.

In short, the gauge group is broken down to
$SU(3)\times SU(2)\times U(1)$ for the zero modes,
and the deconstruction results match the
 continuum results except for $A_5$ and $U_i$.

\section{Higgs Mechanism Deconstruction of
$SU(5)$ Breaking on $M^4\times S^1/(Z_2\times Z_2')$ }
In this section, we would like to discuss the non-supersymmetric
$SU(5)$ breaking on the space-time $M^4\times S^1/(Z_2\times Z_2')$. 
By the way, the $SU(5)$ breaking on the space-time 
$M^4\times S^1/Z_2$ can be discussed similarly.

\subsection{$SU(5)$ Breaking on $M^4\times S^1/(Z_2\times Z_2')$}
Our convention is similar to that in the subsection 2.1.
The orbifold $S^1/(Z_2\times Z_2')$ is obtained by $S^1$ moduloing
the following equivalent classes:
\begin{equation}
 y\sim -y~,~y'\sim -y'~,~\,
\end{equation}
where $y'$ is defined as $y' \equiv y-\pi R/2$. 

For a generic bulk multiplet $\Phi(x^{\mu}, y)$ which fills a 
representation of the gauge
group $G$,
we can define two parity operators $P$ and $P'$ for the $Z_2$ and
$Z_2'$ symmetries, respectively
\begin{eqnarray}
\Phi(x^{\mu},y)&\to \Phi(x^{\mu},-y )=\eta_{\Phi} 
P^{\l_{\Phi}}\Phi(x^{\mu},y)
(P^{-1})^{m_{\Phi}}~,~\,
\end{eqnarray}
\begin{eqnarray}
\Phi(x^{\mu},y')&\to \Phi(x^{\mu},-y' )= \eta'_{\Phi} (P')^{\l_{\Phi}} 
\Phi(x^{\mu},y')
(P^{'-1})^{m_{\Phi}}
 ~,~\,
\end{eqnarray}
where $\eta_{\Phi} = \pm 1$ and $\eta'_{\Phi} = \pm 1$.

Denoting the field $\phi$ with ($P$, $P'$)=($\pm, \pm$)
by $\phi_{\pm \pm}$, we obtain the KK mode expansions
\begin{eqnarray}
  \phi_{++} (x^\mu, y) &=& 
      \sum_{n=0}^{\infty} \frac{1}{\sqrt{2^{\delta_{n,0}} \pi R}} 
      \phi^{(2n)}_{++}(x^\mu) \cos{2ny \over R}~,~\,
\end{eqnarray}
\begin{eqnarray}
  \phi_{+-} (x^\mu, y) &=& 
      \sum_{n=0}^{\infty} \frac{1}{\sqrt{\pi R}} 
      \phi^{(2n+1)}_{+-}(x^\mu) \cos{(2n+1)y \over R}~,~\,
\end{eqnarray}
\begin{eqnarray}
  \phi_{-+} (x^\mu, y) &=& 
      \sum_{n=0}^{\infty} \frac{1}{\sqrt{\pi R}} \,
      \phi^{(2n+1)}_{-+}(x^\mu) \sin{(2n+1)y \over R}~,~\,
\end{eqnarray}
\begin{eqnarray}
  \phi_{--} (x^\mu, y) &=& 
      \sum_{n=0}^{\infty} \frac{1}{\sqrt{\pi R}} 
      \phi^{(2n+2)}_{--}(x^\mu) \sin{(2n+2)y \over R}~.~\,
\end{eqnarray}
The 4-dimensional fields $\phi^{(2n)}_{++}$, $\phi^{(2n+1)}_{+-}$, 
$\phi^{(2n+1)}_{-+}$ and $\phi^{(2n+2)}_{--}$ acquire masses 
$2n/R$, $(2n+1)/R$, $(2n+1)/R$ and $(2n+2)/R$ upon the compactification.
Zero modes are contained only in $\phi_{++}$ fields,
thus, the matter content of massless sector is smaller
than that of the full 5-dimensional multiplet.
Moreover, only $\phi_{++}$ and $\phi_{+-}$ fields have non-zero
values at $y=0$, and only $\phi_{++}$ and $\phi_{-+}$
 fields have non-zero values at $y=\pi R/2$.

Under parity $P$,
the gauge fields $A_M$ transform as 
\begin{eqnarray}
  A_{\mu} (x^\mu, -y) &=& P A_{\mu} (x^\mu, y) P^{-1}
~,~\,
\end{eqnarray}
\begin{eqnarray}
  A_{5} (x^\mu, -y) &=& - P A_{5} (x^\mu, y) P^{-1}
~.~\,
\end{eqnarray}
And under parity $P'$, the gauge field transformations
are similar to those under $P$. 

We choose the following matrix representations for parity operators 
$P$ and $P'$ that are expressed in the adjoint representation of SU(5)
\begin{equation}
P={\rm diag}(+1, +1, +1, +1, +1)
~,~ P'={\rm diag}(-1, -1, -1, +1, +1)~.~\,
\end{equation}
So, under $P'$ parity,
the $SU(5)$ gauge generators $T^A$ where A=1, 2, ..., 24 for SU(5)
are separated into two sets: $T^a$ are the gauge generators for
the Standard Model gauge group, and $T^{\hat a}$ are the other broken
gauge generators
\begin{equation}
P~T^a~P^{-1}= T^a ~,~ P~T^{\hat a}~P^{-1}= T^{\hat a}
~,~\,
\end{equation}
\begin{equation}
P'~T^a~(P')^{-1}= T^a ~,~ P'~T^{\hat a}
~(P')^{-1}= - T^{\hat a}
~.~\,
\end{equation}

Therefore, the masses for gauge fields
$A_{\mu}^a$, $A_{\mu}^{\hat a}$, $A_5^a$ and
$A_5^{\hat a}$ are $2n/R$, $(2n+1)/R$, $(2n+2)/R$
and $(2n+1)/R$, respectively.  For the zero modes,
the gauge group is $SU(3)\times SU(2)\times U(1)$. 
Including the KK modes, the gauge groups at
$y=0$ and $y=\pi R/2$ are $SU(5)$ and 
$SU(3)\times SU(2)\times U(1)$, respectively.

Moreover, assuming that there exists a pair of Higgs 5-plets $H_u$ and 
$H_d$ in the 
bulk and  $\eta_{H_u}=\eta_{H_d}=+1$,  we obtain that for each 5-plet, the
doublet mass is $2n/R$, and the triplet mass is $(2n+1)/R$. So, we solve
the doublet-triplet splitting problem.
The parities and masses of the fields
in the $SU(5)$ gauge and Higgs multiplets are given in Table 2.

\renewcommand{\arraystretch}{1.4}
\begin{table}[t]
\caption{Parity assignment and masses of the fields ($n\ge 0$) in the 
$SU(5)$ 
 gauge and Higgs multiplets for the model with
 $SU(5)$ breaking on $M^4\times S^1/(Z_2\times Z_2')$.
\label{tab:chiral}}
\vspace{0.4cm}
\begin{center}
\begin{tabular}{|c|c|c|}
\hline        
$(P,P')$ & field & mass ($n=0,1, 2, ...$)\\ 
\hline
$(+,+)$ &  $A^a_{\mu}$, $H^D_u$, $H^D_d$ & $2n/R$ \\
\hline
$(+,-)$ &  $A^{\hat{a}}_{\mu}$,  $H^T_u$, $H^T_d$ & $(2n+1)/R$ \\
\hline
$(-,+)$ &  $A^{\hat{a}}_5$ & $(2n+1)/R$ \\
\hline
$(-,-)$ &  $A^{a}_5$ & $(2n+2)/R$  \\
\hline
\end{tabular}
\end{center}
\end{table}

\subsection{Higgs Mechanism Deconstruction}
We consider the $SU(5)^{N}\times (SU(3)\times SU(2)\times U(1))$
 gauge theory with bifundamental
fields $U_i$ ($i=0, 1, ..., N-1$), $U_c$ and $U_w$ as follows
\begin{equation}
\label{simpless}
\begin{array}{c|cccccccr}
     &SU(5)_0 & SU(5)_1 & SU(5)_2 & \cdots & SU(5)_{N-1} & SU(3) & SU(2) & 
U(1)_Y  \\
\hline
{\vrule height 15pt depth 5pt width 0pt}
U_0 & \Yfund  & \overline{\Yfund}  & 1  & \cdots & 1 &1 & 1 & 0\\

U_1 &1 & \Yfund  & \overline{\Yfund} & \cdots & 1 & 1 & 1 & 0\\
U_2  & 1  &1     & \Yfund  & \cdots & 1 & 1 & 1 & 0 \\
  \vdots & \vdots & \vdots & \vdots & \ddots & \vdots & \vdots & \vdots& 
\vdots \\
U_{c} &1  & 1 & 1 & \cdots & \Yfund & \overline{\Yfund} & 1 &\frac{1}{3}\\
U_{w} &1  & 1 & 1 & \cdots & \Yfund & 1 & \Yfund & -\frac{1}{2}\\
\end{array}. \nonumber
\end{equation}
This kind of models has been discussed recently in Ref~\cite{CKT, NW}.

The effective action is
\begin{eqnarray}
 S&=&\int d^4x \sum_{i=0}^{N-2} \left(-{1\over {4 g^2}} Tr F_i^2 + 
Tr[(D_{\mu} U_i)^{\dagger} D^{\mu} U_i]+...\right)
\nonumber\\&&
-{1\over {4 g^2}} Tr F_c^2 + Tr[(D_{\mu} U_c)^{\dagger} D^{\mu} U_c]+...
\nonumber\\&&
-{1\over {4 g^2}} Tr F_w^2 + Tr[(D_{\mu} U_w)^{\dagger} D^{\mu} U_w]+...
~.~\,
\end{eqnarray}
Because we consider the Higgs mechanism deconstruction of $SU(5)$ breaking
on the space-time $M^4\times S^1/(Z_2\times Z_2')$, at GUT scale, we 
should
 take the same gauge couplings for
all the gauge groups $SU(5)^{N}\times (SU(3)\times SU(2)\times U(1))$. 

And we choose the following
VEVs for $U_i$, $U_c$ and $U_w$
\begin{eqnarray}
 < U_i>
&=& {\rm diag} (v/\sqrt 2, v/\sqrt 2, v/\sqrt 2, v/\sqrt 2, v/\sqrt 2),
~{\rm for }~i=0, 1, 2, ..., N-1 
~,~\,
\end{eqnarray}
\begin{equation}
 < U_c> = {\rm diag} (v/\sqrt 2, v/\sqrt 2, v/\sqrt 2) ~,~
< U_w> = {\rm diag} (v/\sqrt 2, v/\sqrt 2) 
~.~\,
\end{equation}
 
The $(N+1)\times (N+1)$ mass matrix for the
Standard Model gauge boson is 
\begin{equation}
\label{SM32}
M_{SM}^2 = g^2 v^2\left(
\begin{array}{ccccccc}
1&-1&0&\cdots&0 & 0 &0\\
-1&2&-1& \cdots&0 & 0&0\\
0&-1 &2 &\cdots&0 & 0&0\\
& & & \ddots & & &\\
0&0&0&\cdots&-1&2&-1\\
0&0&0&\cdots&0&-1&1
\end{array} \right).
\end{equation}
However, the mass spectrum and eigenvector of above mass matrix
 do not exactly match those of $A_{\mu}^a$ in the
 last subsection because of the fixed
point. The subtle point is similar to that in the brane models
on $M^4\times S^1/Z_2$ or $M^4\times R^1/Z_2$ where the brane
tension at fixed point on the quotient space is half of
that on the covering space~\cite{BT}.
And on the covering space $S^1$, the mass matrix for the 
Standard Model gauge boson
is similar to that in Eq. (\ref{SM22}), then
the continuum results and the deconstruction results do exactly
match. In short, the $n-th$ column eigenvector $\alpha_n$ and eigenvalue
$M_n^2$ should
satisfy the following equation
\begin{eqnarray}
(M_{SM}^2)_{ij} \alpha_n^j= \left[1- {1\over 2} 
(\delta_{0i}+\delta_{Ni})\right]
 M_n^2 \alpha_n^i 
~.~\,
\end{eqnarray}
The mass spectrum or eigenvalue is 
\begin{eqnarray}
M_n^2= 4 g^2 v^2 \sin^2({{n\pi}\over {2N}})
~,~\,
\end{eqnarray}
where $n=0, 1, 2, ..., N$.
And the corresponding $n-th$ eigenvector is
\begin{eqnarray}
\alpha_n=(\alpha_{n}^0, \alpha_n^1, ..., \alpha_n^N)
~,~\,
\end{eqnarray}
where
\begin{eqnarray}
 \alpha_n^j &=& {{\sqrt 2}\over {\sqrt {N+1}}} 
\cos({{2 j n\pi}\over {2N}}), ~j=0, 1, ..., N
~.~\,
\end{eqnarray}
Noticing that $R=2N/(gv)$, we obtain that the $n-th$ eigenvector matches
the 5-dimensional wave function ($\cos(2ny/R)$) of 
the $n-th$ KK mode for $A_{\mu}^a$ in the last 
subsection ($\phi_{++}$),
and for $n<< N$, the $n-th$ mass (eigenvalue) matches the mass of the
$n-th$ KK mode for $A_{\mu}^a$ in the last subsection.

The $N\times N$ mass matrix for the
non-Standard Model gauge boson is 
\begin{equation}
\label{NSM32}
M_{NSM}^2 = g^2 v^2\left(
\begin{array}{ccccccc}
1&-1&0&\cdots&0 & 0 &0\\
-1&2&-1& \cdots&0 & 0&0\\
0&-1 &2 &\cdots&0 & 0&0\\
& & & \ddots & & &\\
0&0&0&\cdots&-1&2&-1\\
0&0&0&\cdots&0&-1&2
\end{array} \right).
\end{equation}
Similarly, the mass spectrum and eigenvector
of above matrix do not match those of $A_{\mu}^{\hat a}$
in the last subsection due to the fixed
point. On the covering space $S^1$, the mass matrix
is reducible and the irreducible mass matrix is similar to that
in Eq. (\ref{A532}). Thus, the continuum results and the deconstruction 
results
do exactly match on the covering space $S^1$.
In short, the $n-th$ eigenvector $\alpha_n$ and eigenvalue
$M_n^2$ should
satisfy the following equation
\begin{eqnarray}
(M_{NSM}^2)_{ij} \alpha_n^j= \left[1-{1\over 2} \delta_{0i}\right]
 M_n^2 \alpha_n^i 
~,~\,
\end{eqnarray}
where $j=0, 1, 2, ..., N-1$.
The mass spectrum or eigenvalue is 
\begin{eqnarray}
M_n^2= 4 g^2 v^2 \sin^2({{(2n+1)\pi}\over {4N}})
~,~\,
\end{eqnarray}
where $n=0, 1, 2, ..., N-1$.
And the corresponding $n-th$ eigenvector is
\begin{eqnarray}
\alpha_n=(\alpha_{n}^0, \alpha_n^1, ..., \alpha_n^{N-1})
~,~\,
\end{eqnarray}
where
\begin{eqnarray}
 \alpha_n^j &=& {{\sqrt 2}\over {\sqrt {N}}} 
\cos({{j (2n+1)\pi}\over {2N}}), ~j=0, 1, ..., N-1
~.~\,
\end{eqnarray}
Noticing that $R=2N/(gv)$, we obtain that the $n-th$ eigenvector matches
the 5-dimensional wave function ($\cos((2n+1)y/R)$) of
the $n-th$ KK mode for $A_{\mu}^{\hat a}$ in the 
last subsection ($\phi_{+-}$),
and for $n<< N$, the $n-th$ mass (eigenvalue) matches the mass of the
$n-th$ KK mode for $A_{\mu}^{\hat a}$ in the last subsection.

In addition,
we would like to discuss the deconstruction of 5-dimensional
fields $\phi_{-+}$ and $\phi_{--}$ because we will have this kind
of field expansions when we discuss the effective deconstruction
 scenario in the next section. So, let us give the mass matrix,
 eigenvalues and eigenvectors here. 

The $N\times N$  mass matrix for the deconstruction of 5-dimensional
field $\phi_{-+}$  is 
\begin{equation}
\label{mat1}
M_{-+}^2 = g^2 v^2\left(
\begin{array}{ccccccc}
2&-1&0&\cdots&0 & 0 &0\\
-1&2&-1& \cdots&0 & 0&0\\
0&-1 &2 &\cdots&0 & 0&0\\
& & & \ddots & & &\\
0&0&0&\cdots&-1&2&-1\\
0&0&0&\cdots&0&-1&1
\end{array} \right).
\end{equation}
The mass spectrum or eigenvalue is 
\begin{eqnarray}
M_n^2= 4 g^2 v^2 \sin^2({{(2n+1)\pi}\over {4N}})
~,~\,
\end{eqnarray}
where $n=0, 1, 2, ..., N-1$.
And the corresponding $n-th$ eigenvector is
\begin{eqnarray}
\alpha_n=(\alpha_{n}^1, \alpha_n^2, ..., \alpha_n^N)
~,~\,
\end{eqnarray}
where
\begin{eqnarray}
 \alpha_n^j &=& {{\sqrt 2}\over {\sqrt {N}}} 
\sin({{j (2n+1)\pi}\over {2N}}), ~j=1, 2, ..., N
~.~\,
\end{eqnarray}

The $(N-1)\times (N-1)$  mass matrix for the deconstruction of 
5-dimensional
field $\phi_{--}$  is 
\begin{equation}
\label{A532}
M_{--}^2 = g^2 v^2\left(
\begin{array}{ccccccc}
2&-1&0&\cdots&0 & 0 &0\\
-1&2&-1& \cdots&0 & 0&0\\
0&-1 &2 &\cdots&0 & 0&0\\
& & & \ddots & & &\\
0&0&0&\cdots&-1&2&-1\\
0&0&0&\cdots&0&-1&2
\end{array} \right).
\end{equation}
Because the
field $\phi_{--}$ vanishes at the fixed point, 
we would like to emphasize that the mass spectrum
and eigenvector for the physical field are the eigenvalue and eigenvector
of above mass matrix.
The mass spectrum or eigenvalue is 
\begin{eqnarray}
M_n^2= 4 g^2 v^2 \sin^2({{(n+1)\pi}\over {2N}})
~,~\,
\end{eqnarray}
where $n=0, 1, 2, ..., N-2$.
And the corresponding $n-th$ eigenvector is
\begin{eqnarray}
\alpha_n=(\alpha_{n}^1, \alpha_n^2, ..., \alpha_n^{N-1})
~,~\,
\end{eqnarray}
where
\begin{eqnarray}
 \alpha_n^j &=& {{\sqrt 2}\over {\sqrt {N}}} 
\sin({{j (n+1)\pi}\over {N}}), ~j=1, 2, ..., N-1
~.~\,
\end{eqnarray}

Moreover, we can discuss the deconstruction of Higgs fields $H_u$ and 
$H_d$.
We assume that under the $i-th$ gauge group $SU(5)_i$, there is a pair of 
Higgs
5-plets $H_u^i$ and $H_d^i$ where $i=0, 1, 2, ..., N-1$, and for the 
$N-th$ gauge
group, there is a pair of Higgs doublets $H_u^{ND}$ and $H_d^{ND}$. And we
consider the following potential
\begin{eqnarray}
V&=& 2 g^2 \sum_{i=0}^{N-2} (|U_i (H_u^i)^{\dagger}|^2 + |H_u^{i+1} U_i|^2 
+
|U_i H_d^i|^2 + |(H_d^{i+1})^{\dagger} U_i|^2  )
\nonumber\\&& 
-\sqrt 2 g^2 v \sum_{i=0}^{N-2} (H_u^{i+1} U_i (H_u^i)^{\dagger} + 
(H_d^{i+1})^{\dagger} U_i H_d^i + H. C.)
\nonumber\\&& 
+ 2 g^2  ( |U_w (H_u^{(N-1)D})^{\dagger}|^2 + |H_u^{ND} U_w|^2 +
|U_w H_d^{(N-1)D}|^2 + |(H_d^{ND})^{\dagger} U_w|^2 )
\nonumber\\&& 
-\sqrt 2 g^2 v  (H_u^{ND} U_w (H_u^{(N-1)D})^{\dagger} + 
(H_d^{ND})^{\dagger} U_w H_d^{(N-1)D} + H. C.)
~.~\,
\end{eqnarray}
This potential for a pair of Higgs
5-plets $H_u^i$ and $H_d^i$ comes from the deconstruction of $(D_5 
H_u)^{\dagger} D_5 H_u $ 
and $(D_5 H_d)^{\dagger} D_5 H_d $ in 5-dimensional theory, 
and the coefficients are determined by the normalization
which is compatible with that of gauge fields. So,
we obtain that the mass matrix for the doublet $H_u^{iD}$ or $H_d^{iD}$  
is the same as that
for the Standard Model gauge boson in Eq. (\ref{SM32}), and the 
mass matrix for the triplet $H_u^{iT}$ or $H_d^{iT}$  is the same as that
for the non-Standard Model gauge boson in Eq. (\ref{NSM32}).
Therefore, similar to the discussions
for $A_{\mu}^a$ and $A_{\mu}^{\hat a}$, the deconstruction results match 
the
 continuum results in Table 2 exactly.
And then, we solve the doublet-triplet splitting problem.

Furthermore, by counting the number of massless modes,
 we can prove that the bifundamental fields $U_i$ can not match
the gauge fields $A_5$ because $U_i$s are Higgs fields. The correct
deconstruction of the $A_5$ fields should have 0 massless modes, but, the
$U_i$s field will have at least $24N$ massless modes that are the
Goldstone bosons and give 
 masses to the longitudinal components of massive gauge bosons.

In short, the gauge group is broken down to
$SU(3)\times SU(2)\times U(1)$ for the zero modes,
and the deconstruction results match the
 continuum results except for $A_5$ and $U_i$.

\section{Gauge Symmetry Breaking by the Discrete Symmetry on the Theory 
Space
in the Effective Deconstruction Scenario}
It seems to us that 
the Higgs mechanism deconstructions of the 
gauge symmetry breaking by discrete symmetry on
the space-time $M^4\times S^1$ or 
$M^4\times S^1/(Z_2\times Z_2')$ in the last two sections
 might not be the real deconstructions. The key point is that,
the bifundamental field $U_i$, which is the Schwinger line integral
along the fifth dimension,
should be considered as the field $A_5$.
However, the mass spectrum and the 5-dimensional wave function for
the KK modes of $A_5$ 
 can not match those for $U_i$ 
in the Higgs mechanism deconstruction scenario in above two sections by
counting the massless modes.
In order to have exact match, we propose the effective
deconstruction scenario and discuss the gauge symmetry breaking by 
the discrete symmetry on theory space.

Before we propose the effective
deconstruction scenario, 
let us consider the 5-dimensional $SU(5)$ theory with
a pair of Higgs 5-plets $H_u$ and $H_d$ on the space-time $M^4\times S^1$.
 If we latticized the fifth dimension with $N+1$ sites,
we obtain the following mass terms for the fields $A_M^i$, $H_u^i$, 
$H_d^i$
from the latticized kinetic terms of the fields along the fifth dimension, 
{\it i. e.},
$\partial_5 \phi \partial^5 \phi$ for a generic bulk field $\phi (x^{\mu}, 
y)$
\begin{eqnarray}
\label{EDS}
V &=& \sum_{i=0}^{N-1} \left({{N+1}\over {2 \pi R}}\right)^2
\left( {1\over 2} (A_M^{i+1}-A_M^i)^2 + |H_u^{i+1}-H_u^i|^2
+ |H_d^{i+1}-H_d^i|^2 \right)
\nonumber\\&&
+ \left({{N+1}\over {2 \pi R}}\right)^2
\left({1\over 2} (A_M^{N}- \Gamma A_M^{0} \Gamma^{-1})^2 
+ |H_u^{N}- \eta_{H_u} \Gamma H_u^i|^2
\right.\nonumber\\&&\left.
+ |H_d^{i+1}- \eta_{H_d} \Gamma^{-1} H_d^i|^2\right)
~,~\,
\end{eqnarray}
where the subscript $M$ denotes $0, 1, 2, 3, 5$ ($\mu$ and $5$), 
 $\Gamma$ is a $5\times 5$ matrix and is a generator
of $Z_n$ group, which is a subgroup of $\pi_1(S^1)=Z$, {\it i. e.},
$\Gamma^n=1$. Of course, there exist some
other terms, but we are not interested in them here.
It is not hard for one to prove that the 5-dimensional $SU(5)$ gauge
theory on $M^4\times S^1$ is equivalent to the 4-dimensional $SU(5)^{N+1}$
gauge theory with above effective potential and other terms in the large 
$N$
limit, {\it i. e.}, $N \rightarrow +\infty$.
In addition, we would like to point out that the $SU(5)^{N+1}$ gauge
theory with above effective potential preserve only the $SU(5)/\Gamma$ 
gauge
symmetry, where the gauge group $SU(5)/\Gamma$ is the commutant of 
$\Gamma$ in $SU(5)$,
mathematically speaking, 
\begin{equation}
SU(5)/\Gamma\equiv \{g \in SU(5)|g\Gamma=\Gamma g\}~.~\,
\end{equation}
Of course, the Lagrangian is not $SU(5)^{N+1}$ gauge invariant, and then, 
the
theory is non-renormalizable. However, this latticized theory is correct 
because
from the point of view of 4-dimensional effective theory,
the 5-dimensional $SU(5)$ gauge
theory on $M^4\times S^1$ preserve only the $SU(5)/\Gamma$ gauge symmetry 
for the zero modes,
the gauge symmetries for the non-zero KK modes of 5-dimensional gauge 
fields are completely broken,
and the 5-dimensional theory is non-renormalizable.

Therefore, we can consider the $SU(5)^{N+1}$ gauge theory where in 
particular
the gauge fields $A_5$ or the corresponding link fields $U_i$ do not have 
the VEVs,
 and we introduce above effective potential for the mass terms
 by hand. In this approach, we can consider the
 discrete symmetry on
 theory space and discuss the gauge symmetry breaking.  Moreover,
the continuum results exactly match the deconstruction results.
Because we add the mass terms by hand and our theory is an effective 
theory, we call this deconstruction scenario as 
effective deconstruction scenario.
However, we would like to emphasize that the effective deconstruction 
scenario is 
the non-renormalizable theory, and how to construct a renomalizable
effective deconstruction scenario deserves further study and is out of the 
scope
of this paper.

\subsection{ Deconstruction of $SU(5)$ Breaking on $M^4\times S^1$ by 
Wilson Line}
Considering the $SU(5)^{N+1}$ gauge theory with
bifundamental link fields $U_i$ given in Eq. (\ref{wilson}) in subsection 
2.2, and assuming that 
there exists one pair
of Higgs 5-plets $H_u^i$ and $H_d^i$ under the gauge group $SU(5)_i$ where 
$i=0, 1, 2, ..., N$,
 we choose the following effective potential
\begin{eqnarray}
\label{EDS1}
V &=& \sum_{i=0}^{N-1} \left({{N+1}\over {2 \pi R}}\right)^2
\left( {1\over 2} (A_{\mu}^{i+1}-A_{\mu}^i)^2 + {1\over 2} (U_{i+1}-U_i)^2
\right.\nonumber\\&&\left.
+ |H_u^{i+1}-H_u^i|^2
+ |H_d^{i+1}-H_d^i|^2 \right)
\nonumber\\&&
+ \left({{N+1}\over {2 \pi R}}\right)^2
\left({1\over 2} (A_{\mu}^{N}- \Gamma A_{\mu}^{0} \Gamma^{-1})^2 
+ {1\over 2} (U_{N}- \Gamma U_{0} \Gamma^{-1})^2
\right.\nonumber\\&&\left.
+ |H_u^{N}- \eta_{H_u} \Gamma H_u^i|^2
+ |H_d^{i+1}- \eta_{H_d} \Gamma^{-1} H_d^i|^2\right)
~,~\,
\end{eqnarray}
where
\begin{eqnarray}
\Gamma = {\rm diag} (-1, -1, -1, +1, +1) ~,~
\end{eqnarray}
and  $\eta_{H_u}=\eta_{H_d}=+1$. So, we obtain that 
the parities of all the fields 
are the same as those in the subsection 2.1 (see Table 1), and
the deconstruction results exactly match the 
continuum results.

\subsection{ Deconstruction of $SU(5)$ Breaking on $M^4\times 
S^1/(Z_2\times Z_2')$}
We consider the $SU(5)^{4N}$
 gauge theory with bifundamental
fields $U_i$ as follows\footnote{For convenience to
define $Z_2$ and $Z_2'$ symmetries, we define half
$U_i$ as $U_i^{\dagger}$ in the previous sections. Because we consider
the non-supersymmetric theory, there is no anomaly problem.}

\begin{equation}
\label{orbifold}
\begin{array}{c|cccccc}
      & SU(5)_0 & SU(5)_1 & SU(5)_2 & \cdots & SU(5)_{N-1} & SU(5)_N \\
\hline
{\vrule height 15pt depth 5pt width 0pt}
 U_0       & \Yfund            & \overline{\Yfund}  & 1     & \cdots & 1 
&1 \\
  U_1       & 1       & \Yfund & \overline{\Yfund}   & \cdots & 1 & 1 \\
  \vdots    & \vdots & \vdots & \ddots & \vdots & \vdots \\
  U_{N-1}   & 1 & 1 & 1& \cdots & \Yfund & \overline{\Yfund} \\
\end{array}. \nonumber
\end{equation}

\begin{equation}
\label{orbifold}
\begin{array}{c|cccccc}
      & SU(5)_N & SU(5)_{N+1} & SU(5)_{N+2} & \cdots & SU(5)_{2N-1} & 
SU(5)_{2N} \\
\hline
{\vrule height 15pt depth 5pt width 0pt}
 U_N       & \overline{\Yfund}            & \Yfund  & 1     & \cdots & 1 
&1 \\
  U_{N+1}       & 1       & \overline{\Yfund} & \Yfund   & \cdots & 1 & 1 
\\
  \vdots    & \vdots & \vdots & \ddots & \vdots & \vdots \\
  U_{2N-1}   & 1 & 1 & 1& \cdots & \overline{\Yfund} & \Yfund \\
\end{array}. \nonumber
\end{equation}

\begin{equation}
\label{orbifold}
\begin{array}{c|cccccc}
      & SU(5)_{2N} & SU(5)_{2N+1} & SU(5)_{2N+2} & \cdots & SU(5)_{3N-1} & 
SU(5)_{3N} \\
\hline
{\vrule height 15pt depth 5pt width 0pt}
 U_{2N}       & \Yfund            & \overline{\Yfund}  & 1     & \cdots & 
1 &1 \\
  U_{2N+1}       & 1       & \Yfund & \overline{\Yfund}   & \cdots & 1 & 1 
\\
  \vdots    & \vdots & \vdots & \ddots & \vdots & \vdots \\
  U_{3N-1}   & 1 & 1 & 1& \cdots & \Yfund & \overline{\Yfund} \\
\end{array}. \nonumber
\end{equation}

\begin{equation}
\label{orbifold}
\begin{array}{c|cccccc}
      & SU(5)_{3N} & SU(5)_{3N+1} & SU(5)_{3N+2} & \cdots & SU(5)_{4N-1} & 
SU(5)_{0} \\
\hline
{\vrule height 15pt depth 5pt width 0pt}
 U_{3N}       & \overline{\Yfund}            & \Yfund  & 1     & \cdots & 
1 &1 \\
  U_{3N+1}       & 1       & \overline{\Yfund} & \Yfund   & \cdots & 1 & 1 
\\
  \vdots    & \vdots & \vdots & \ddots & \vdots & \vdots \\
  U_{4N-1}   & 1 & 1 & 1& \cdots & \overline{\Yfund} & \Yfund \\
\end{array}. \nonumber
\end{equation}

On the theory space, there are $4N$ sites and $4N$ bifundamental link 
fields.
And we would like to add one pair of Higgs 5-plets $H_u^i$ and $H_d^i$ on 
the $i-th$ site.
Moreover, we introduce the following effective potential for the mass 
terms
\begin{eqnarray}
V &=& \sum_{i=0}^{4N-1} \left({{2N}\over { \pi R}}\right)^2
\left( {1\over 2} (A_{\mu}^{i+1}-A_{\mu}^i)^2 +
{1\over 2} (U_i-U_{i+1})^2
\right.\nonumber\\&&\left.
+ |H_u^{i+1}-H_u^i|^2
+ |H_d^{i+1}-H_d^i|^2 \right)
~.~\,
\end{eqnarray}
So, the mass matrix for each field is similar to that in Eq. (\ref{SM22}).

Now, let us discuss the discrete symmetries on the theory space.
We define that $i'\equiv i+3N$ for $0 \leq i < N$ and
$i'\equiv i-N$ for $N \leq i < 4N$. The $Z_2$ and $Z_2'$ symmetries
on theory space are defined by the following equivalent classes
\begin{equation}
 i \sim 4N-i ~{\rm for}~ Z_2~,~i'\sim 4N-i'~{\rm for}~ Z_2'~.~\,
\end{equation}

For a generic multiplet $\Phi^i (x^{\mu})$ ($i=0, 1, 2, ..., 4N$)
 which fills a representation of the gauge
group $SU(5)_i$,
we can define two parity operators $P$ and $P'$ for the $Z_2$ and
$Z_2'$ symmetries, respectively
\begin{eqnarray}
\Phi^i(x^{\mu})&\to \Phi^{4N-i}(x^{\mu})=\eta_{\Phi} 
P^{\l_{\Phi}}\Phi^i(x^{\mu})
(P^{-1})^{m_{\Phi}}~,~\,
\end{eqnarray}
\begin{eqnarray}
\Phi^{i'}(x^{\mu})&\to \Phi^{4N-i'}(x^{\mu} )= \eta'_{\Phi} 
(P')^{\l_{\Phi}} \Phi^{i'}(x^{\mu})
(P^{'-1})^{m_{\Phi}}
 ~,~\,
\end{eqnarray}
where $\eta_{\Phi} = \pm 1$ and $\eta'_{\Phi} = \pm 1$.

Denoting the physical field $\tilde \phi$ with ($P$, $P'$)=($\pm, \pm$)
by ${\tilde \phi}_{\pm \pm}$, we obtained the physical fields
${\tilde \phi}^{(n)}_{\pm \pm} (x^{\mu})$ expansion in terms of
the site fields $\phi^j (x^\mu)$
\begin{eqnarray}
 {\tilde \phi}^{(2n)}_{++} (x^\mu) &=& {1\over {\sqrt {2N}}} 
\sum_{j=0}^{4N-1}
\cos({{j2n\pi}\over 2N}) \phi^j (x^\mu) ~,~\,
\end{eqnarray}
\begin{eqnarray}
 {\tilde \phi}^{(2n+1)}_{+-} (x^\mu) &=& {1\over {\sqrt {2N}}} 
\sum_{j=0}^{4N-1}
\cos({{j(2n+1)\pi}\over {2N}}) \phi^j (x^\mu) ~,~\,  
\end{eqnarray}
\begin{eqnarray}
 {\tilde \phi}^{(2n+1)}_{-+} (x^\mu) &=& {1\over {\sqrt {2N}}} 
\sum_{j=0}^{4N-1}
\sin({{j(2n+1)\pi}\over {2N}}) \phi^j (x^\mu) ~,~\, 
\end{eqnarray}
\begin{eqnarray}
{\tilde \phi}^{(2n+2)}_{--} (x^\mu) &=& {1\over {\sqrt {2N}}} 
\sum_{j=0}^{4N-1}
\sin({{j(2n+2)\pi}\over {2N}}) \phi^j (x^\mu) ~,~\,
\end{eqnarray}
The physical fields ${\tilde \phi}^{(2n)}_{++}$, ${\tilde 
\phi}^{(2n+1)}_{+-}$, 
${\tilde \phi}^{(2n+1)}_{-+}$ and ${\tilde \phi}^{(2n+2)}_{--}$ acquire 
masses 
${{4N}\over {\pi R}} \sin({{2n\pi}\over 4N})$,
 ${{4N}\over {\pi R}} \sin({{(2n+1)\pi}\over 4N})$,
${{4N}\over {\pi R}} \sin({{(2n+1)\pi}\over 4N})$ and
${{4N}\over {\pi R}} \sin({{(2n+2)\pi}\over 4N})$, respectively,
where $n=0, 1, 2$, ..., $2N-1$.
In the large $N$ limit, or $N\rightarrow +\infty$ or $n < < N$, the 
physical fields ${\tilde \phi}^{(2n)}_{++}$, ${\tilde 
\phi}^{(2n+1)}_{+-}$, 
${\tilde \phi}^{(2n+1)}_{-+}$ and ${\tilde \phi}^{(2n+2)}_{--}$ acquire 
masses 
$2n/R, (2n+1)/R, (2n+1)/R$ and $(2n+2)/R$, respectively.
Therefore, the deconstruction results exactly match the continuum results.
In addition, zero modes are contained only in ${\tilde \phi}^{(2n)}_{++}$ 
fields,
thus, the matter content of massless sector is smaller
than that of the full multiplets in the theory.
Moreover, only ${\tilde \phi}^{(2n)}_{++}$ and ${\tilde 
\phi}^{(2n+1)}_{+-}$
 fields have non-zero
values at $i=0$ and $i=2N$, and only ${\tilde \phi}^{(2n)}_{++}$ and
 ${\tilde \phi}^{(2n+1)}_{-+}$
 fields have non-zero values at $i=N$ and $i=3N$.

Under $Z_2$ parity $P$, 
the fields $A_{\mu}^i$, $U_i$, $H_u^i$ and $H_d^i$ transform as 
\begin{eqnarray}
  A_{\mu}^{4N-i} (x^\mu) &=& P A^i_{\mu} (x^\mu) P^{-1}
~,~\,
\end{eqnarray}
\begin{eqnarray}
  U^{4N-i} (x^\mu) &=& - P U^i (x^\mu) P^{-1}
~,~\,
\end{eqnarray}
\begin{eqnarray}
  H_u^{4N-i} (x^\mu) &=& P H_u^i (x^\mu) 
~,~\,
\end{eqnarray}
\begin{eqnarray}
  H_d^{4N-i} (x^\mu) &=& P^{-1} H_d^i (x^\mu)
~.~\,
\end{eqnarray}
And under parity $P'$, the gauge field and Higgs field transformations
are similar to those under $P$. 

We choose the following matrix representations for parity operators 
$P$ and $P'$ that are expressed in the adjoint representation of $SU(5)$
\begin{equation}
P={\rm diag}(+1, +1, +1, +1, +1)
~,~ P'={\rm diag}(-1, -1, -1, +1, +1)~,~\,
\end{equation}
and then, we obtain that the parities of all the fields 
are the same as those in the subsection 3.1 (see Table 2), and
the deconstruction results exactly match
the continuum results.

\section{$G^N$ Unification}
 $G^N$ unification has been discussed previously where
each gauge group $G$ is broken by the Higgs fields in its adjoint 
representation~\cite{GNP}, or the
gauge group $G^N$ is broken by introducing more than one link Higgs
field between the first two sites
recently~\cite{CMW}. The doublet-triplet splitting are also considered in
the deconstruction of $SU(5)$ on the disc $D^2$~\cite{EW}. 
In this section, we would like to briefly discuss the $G^N$ unification
where $G^N$ is broken down to 
$SU(3)\times SU(2)\times U(1)^{n-3}$ by the 
bifundamental link fields in which $n$ is
the rank of group $G$.
We shall discuss the scenario with $G=SU(5)$ as an example,
 and similarly, one can discuss
the scenario with $G=SU(6), SO(10), E_6$, etc, for 
$\pi_1(S_1)=Z$~\cite{JUN}.
The discussions for the 
doublet-triplet splitting
are primitive here, and the natural solution to the 
doublet-triplet splitting problem in a supersymmetric scenario will
be presented elsewhere~\cite{TLWL}.

\subsection{$SU(5)_0\times SU(5)_1$}
The set-up is similar to that in the subsection 2.2
for $N=1$. We choose
the VEVs of $U_0$ and $U_1$ as
\begin{eqnarray}
 < U_0>
&=& {\rm diag} (v/\sqrt 2, v/\sqrt 2, v/\sqrt 2, v/\sqrt 2, v/\sqrt 2)
~,~\,
\end{eqnarray}
\begin{eqnarray}
 < U_1>
&=& {\rm diag} (-v/\sqrt 2, -v/\sqrt 2, -v/\sqrt 2, v/\sqrt 2, v/\sqrt 2)
~.~\,
\end{eqnarray}

The mass matrix for the Standard Model gauge boson is
\begin{equation}
\label{GN1}
  M_{SM}^2 =g^2 v^2 \left(\begin{array}{cc}
    2 & -2   \\ 
    -2 & 2  \\ 
  \end{array} \right)~,~\,
\end{equation}
and the mass matrix for the non-Standard Model gauge boson is
\begin{equation}
\label{GN2}
  M_{NSM}^2 =g^2 v^2 \left(\begin{array}{cc}
    2 & 0   \\ 
    0 & 2  \\ 
  \end{array} \right)~.~\,
\end{equation}
It is easy to check that only the Standard Model gauge
 bosons have zero modes
and the non-Standard Model gauge bosons are massive.

Now, let us discuss the doublet-triplet splitting. Suppose that
$H_u^i$ and $H_d^i$ are in the fundamental and anti-fundamental 
representations
 of $SU(5)_i$ respectively, where $i=0, 1$.
With the following potential
\begin{eqnarray}
V&=&2 g^2 (|U_0 (H_u^0)^{\dagger}|^2 + |H_u^{1} U_0|^2 +
|U_0 H_d^0|^2 + |(H_d^{1})^{\dagger} U_0|^2  )
\nonumber\\&&
+ 2 g^2 (|U_1 (H_u^1)^{\dagger}|^2 + |H_u^{0} U_1|^2 +
|U_1 H_d^1|^2 + |(H_d^{0})^{\dagger} U_1|^2  )
\nonumber\\&& 
-\sqrt 2 g^2 v (H_u^{1} U_0 (H_u^0)^{\dagger} + 
(H_d^{1})^{\dagger} U_0 H_d^0 + H. C.)
\nonumber\\&& 
-\sqrt 2 g^2 v (H_u^{0} U_1 (H_u^1)^{\dagger} + 
(H_d^{0})^{\dagger} U_1 H_d^1 + H. C.)
~,~\,
\end{eqnarray}
we obtain that the mass matrix for the doublet $H_u^{iD}$ or $H_d^{iD}$ is 
the same as that
for the Standard Model gauge boson in Eq. (\ref{GN1}), and the 
mass matrix for the triplet $H_u^{iT}$ or $H_d^{iT}$ is the same as that
for the non-Standard Model gauge boson in Eq. (\ref{GN2}).
Therefore, we solve the doublet-triplet splitting problem.

The possible interesting model is that we put three
5-plets ($\bar 5$) fermions under $SU(5)_0$,  and three 10-plets
($10$) fermions under $SU(5)_1$. Then, the proton decay is
suppressed at least at one loop level.
Of course, there might exist 
anomaly in the model. One way to avoid anomaly is that, we consider
 supersymmetry and three link fields that are in the fundamental
and anti-fundamental representations of 
$SU(5)_0$ and $SU(5)_1$, respectively.

\subsection{$SU(5)_0\times SU(5)_1\times SU(5)_2$}
The set-up is similar to that in the subsection 2.2 for $N=2$. However, we 
choose 
the following VEVs of $U_0$, $U_1$ and $U_2$, which can also give the 
correct mass matrices
for gauge boson and Higgs fields
\begin{eqnarray}
 < U_i>
= {\rm diag} (-v/\sqrt 2, -v/\sqrt 2, -v/\sqrt 2, v/\sqrt 2, v/\sqrt 2),
~{\rm for }~i=0, 1, 2
~.~\,
\end{eqnarray}

The mass matrix for the Standard Model gauge boson is
\begin{equation}
\label{GN3}
  M_{SM}^2 =g^2 v^2 \left(\begin{array}{ccc}
    2 & -1 & -1  \\ 
    -1 & 2 & -1 \\ 
    -1 & -1 & 2 \\
  \end{array} \right)~,~\,
\end{equation}
and the mass matrix for the non-Standard Model gauge boson is
\begin{equation}
\label{GN4}
  M_{NSM}^2 =g^2 v^2 \left(\begin{array}{ccc}
    2 & +1 & +1  \\ 
    +1 & 2 & +1 \\ 
    +1 & +1 & 2 \\
  \end{array} \right)~.~\,
\end{equation}
It is easy to check that only the Standard Model gauge bosons have zero 
modes
and the non-Standard Model gauge bosons are massive.

Now, let us discuss the doublet-triplet splitting.
 Suppose that
$H_u^i$ and $H_d^i$ are in the fundamental and anti-fundamental 
representations
 of $SU(5)_i$ respectively, where $i=0, 1, 2$.
With the following potential
\begin{eqnarray}
V&=&2 g^2 (|U_0 (H_u^0)^{\dagger}|^2 + |H_u^{1} U_0|^2 +
|U_0 H_d^0|^2 + |(H_d^{1})^{\dagger} U_0|^2  )
\nonumber\\&&
2 g^2 (|U_1 (H_u^1)^{\dagger}|^2 + |H_u^{2} U_1|^2 +
|U_1 H_d^1|^2 + |(H_d^{2})^{\dagger} U_1|^2  )
\nonumber\\&&
+ 2 g^2 (|U_2 (H_u^2)^{\dagger}|^2 + |H_u^{0} U_2|^2 +
|U_2 H_d^2|^2 + |(H_d^{0})^{\dagger} U_2|^2  )
\nonumber\\&&
- \sqrt 2 g^2 v (H_u^1 U_0 (H_u^0)^{\dagger} + H_u^2 U_1 (H_u^1)^{\dagger}
+ H_u^0 U_2 (H_u^2)^{\dagger}
 + H. C.)
\nonumber\\&&
- \sqrt 2 g^2 v ( (H_d^1)^{\dagger} U_0 H_d^0 + (H_d^2)^{\dagger} U_1 
H_d^1
+ (H_d^0)^{\dagger} U_2 H_d^2
 + H. C.)
~,~\,
\end{eqnarray}
we obtain that the mass matrix for the doublet $H_u^{iD}$ or $H_d^{iD}$ is 
the same as that
for the Standard Model gauge boson in Eq. (\ref{GN3}), and the 
mass matrix for the triplet $H_u^{iT}$ or $H_d^{iT}$ is the same as that
for the non-Standard Model gauge boson in Eq. (\ref{GN4}).
Thus, we solve the doublet-triplet splitting problem.

The possible interesting model is that we put one family
($\bar 5 + 10$) fermions under one $SU(5)$ gauge group, and define the
$Z_3$ symmetry on the model.

\section{Discussion and Conclusion}
An interesting question is how to realize the Yukawa couplings in the 
deconstruction
scenarios. If the 5-dimensional theory is non-supersymmetric, we can 
consider
the 5-dimensional
Yukawa couplings or 3-brane localized Yukawa couplings. In the 
deconstruction
scenarios, we need to introduce the Yukawa couplings on all the sites for 
the first
case. For the second case, we introduce the Yukawa couplings only on the 
corresponding
particular site for that 3-brane.
For example, if the 3-brane localized Yukawa couplings is on the 3-brane 
at $y=0$ in the
5-dimensional theory, we only introduce the Yukawa couplings on the $0-th$ 
site.
In addition, if the 5-dimensional theory is supersymmetric, we can only 
consider the
 Yukawa couplings (superpotential) on the observable 3-brane at the fixed 
point 
because the 5-dimensional $N=1$ 
supersymmetry is 4-dimensional $N=2$ supersymmetry, and then, the bulk 
Yukawa couplings
are forbidden. In the deconstruction scenarios, we introduce the
Yukawa couplings (superpotential)
on the corresponding site for the observable 3-brane at the fixed point.

In this paper, we deconstruct the non-supersymmetric
 $SU(5)$ breaking by discrete symmetry on the space-time $M^4\times S^1$
 and $M^4\times S^1/(Z_2\times Z_2')$ in the
 Higgs mechanism deconstruction
scenario.
And we explain the subtle point on how to exactly
match the continuum results with the latticized results
on the quotient space $S^1/Z_2$ and $S^1/(Z_2\times Z_2')$.
Because it seems to us that
the Higgs mechanism deconstruction
scenario might not be the real  deconstruction,
we propose the effective deconstruction scenario
and discuss the gauge symmetry breaking by the discrete symmetry
on theory space in this approach.
However, for simplicity, we do not consider supersymmetry
and only discuss GUT group $SU(5)$.
So, it is interesting to consider supersymmetry and
discuss other GUT breaking, for instance,
 $SU(6)$, $SO(10)$, and $E_6$, etc.
Moreover, we can study the deconstruction of the extra space
manifolds like the two-torus $T^2$, the disc $D^2$
 and the annulus $A^2$, and discuss the general 
gauge symmetry and supersymmetry breaking by
the discrete symmetry
on the general theory space
in the effective deconstruction scenario. 
It seems to us that we will have 
the link fields and Higgs unification, suppress the proton
decay by R-symmetry, solve the doublet-triplet splitting
problem and $\mu$ problem. 

As an application,
we suggest the $G^N$ unification where $G^N$ is
broken down to $SU(3)\times SU(2)\times U(1)^{n-3}$
by the bifundamental link fields and the doublet-triplet
splitting can be achieved. Furthermore, we can consider
the general link fields, for example,
$(10, \bar {10})$ for $SU(5)_0\times SU(5)_1$.
In short, with the general $G^N$ unification,
we wish we can solve the tough problems
in the traditional 4-dimensional
GUT models. 

\section*{Acknowledgments}
We would like to thank W. Liao for helpful discussions.
This research was supported in part by the U.S.~Department of Energy under
 Grant No.~DOE-EY-76-02-3071.

\end{document}